\documentclass[aps,prd,amssymb,groupedaddress,twocolumn,showpacs]{revtex4}

\usepackage{graphicx}
\usepackage{dcolumn}
\usepackage{amsmath}
\usepackage{bm}
\begin{document}

\title{Classical Origin of the Spin of Relativistic Pointlike Particles\\ and Geometric interpretation of Dirac Solutions}
\author{S.  Savasta and O. Di Stefano}
\affiliation{Dipartimento di Fisica della Materia e
Tecnologie Fisiche Avanzate, Universit\`{a} di Messina Salita
Sperone 31, I-98166 Messina, Italy}

\begin{abstract}
{
Spin of elementary particles is the only kinematic degree of freedom not having classical corre-
spondence. It arises when seeking for the finite-dimensional representations of the Lorentz group,
which is the only symmetry group of relativistic quantum field theory acting on multiple-component
quantum fields non-unitarily. We study linear transformations, acting on the space
of spatial and proper-time velocities rather than on coordinates. While ensuring the relativistic in-
variance, they avoid these two exceptions: they describe the spin degree of freedom of a pointlike
particle yet at a classical level and form a compact group hence with unitary finite-dimensional rep-
resentations. Within this approach changes of the velocity modulus and direction can be accounted
for by rotations of two independent unit vectors. Dirac spinors just provide the quantum description
of these rotations.}
\end{abstract}
\pacs{11.30.-j,11.30.Cp, 11.10.-z,03.65.-w}
\maketitle

\section{Introduction}

Quantum spin differs from the other quantum observables as position, momentum, energy, angular momentum etc.,
for the absence of  classical correspondence. Pauli described it
in his paper on the exclusion principle \cite{Pauli} as a classical nondescribable two-valuedness.
Accordingly in many textbooks about quantum theory, spin is referred as a non-classical degree of freedom.
Moreover spin cannot be regarded as related to some internal symmetry like hypercharge since it originates from spacetime symmetries i.e. from  finite-dimensional representations of the homogeneous Lorentz group.
It is the only kinematic degree of freedom not having classical correspondence, although it is worth mention that  some composite classical dynamical models with additional variables are able to reproduce after quantization the Dirac electron theory \cite{CS1,CS2}.
More recently a geometric origin of the spin angular momentum has been suggested \cite{Newman}.
The fact that the relativistic Dirac theory automatically includes the effects of spin leads to the conclusion that spin is a quantum relativistic effect. Nevertheless this conclusion is not generally accepted. Weinberg (Ref.\ \cite{weinberg} Chapter 1) wrote: {\em \dots it is difficult to agree that there is anything fundamentally wrong with the relativistic equation for zero spin that forced the development of the Dirac equation -- the problem simply is that the electron happens to have spin $\hbar /2$, not zero.}

Historically, Paul Dirac found the Klein-Gordon equation physically unsatisfactory \cite{Dirac}, thus he  seeked for a relativistically invariant wave equation of first order in time satisfying a Schr\"{o}dinger-like wave equation of the form
\begin{equation}
	i\partial_t \psi = 	\hat H \psi\, .
\label{Sl}\end{equation}
In order to have a   more symmetric relativistic wave equation in the 4-momentum components, Dirac seeked for an equation that because is linear in the time-derivative, it is also linear in space-derivatives, so that $\hat H$ takes the form,
\begin{equation}
	\hat H =  {\bm \alpha} \cdot {\bf \hat p} +  \beta m \, ,
\label{DH}\end{equation}
with ${\bm \alpha}$ and $\beta$ being  independent on spacetime and 4-momentum.
The condition that  Eq.\ (\ref{DH}) provides the correct relationship between energy and momentum
\begin{equation}
	E^2 = {\bf p}^2+m^2\, ,
\label{Ec}\end{equation}
requires that ${\bm \alpha}$ and $\alpha_4\equiv \beta$ obey the  anticommutation rules $\left \{ \alpha_i, \alpha_j \right \}= 2 \delta_{ij}$ ($i=1,4$). Dirac found that a set of $4 \times 4$ matrices satisfying this relation provides the lowest order representation of the four $\alpha_i$. They can be expressed in terms of Pauli matrices $\rho_i$ and $\sigma_i$ belonging to two different Hilbert spaces: ${\alpha}_i = \rho_1 \sigma_i$ ($i=1,3$) and $\beta=\rho_3$. Inserting Eq.\ (\ref{DH}) into (\ref{Sl}) the Dirac equation involving a four-component wavefunction is obtained.

Richard Feynman in his Nobel lecture wrote: {\em Dirac obtained his equation for the
description of the electron by an almost purely mathematical proposition. A
simple physical view by which all the contents of this equation can be seen is
still lacking}.
Dirac's Hamiltonian  seems not to  have a direct correspondence with the classical relativistic Hamiltonian of the free pointlike particle 
\begin{equation}
	H = \sqrt{{\bf p}^2+m^2}\, ,
\label{Ecsr}\end{equation}
in contrast to the nonrelativistic Schr\"{o}dinger-like wave equation which can be derived directly from the Hamiltonian after the quantum operator replacement. Analogously the Klein-Gordon equation can be derived directly from the relativistic relationship (\ref{Ec}).

Actually, in 1928, (the same year of the  Dirac's exceptional achievement) Breit \cite{Breit} provided the lacking correspondent principle, recalling that another way of writing Eq.\ (\ref{Ecsr}) is
\begin{equation}
H = {\bf \dot x} \cdot {\bf p}  + m\sqrt{1- \left| {\bf \dot x} \right|^2}\, \, ,
\label{H_Br}\end{equation}
which is the form at which one usually arrives first in the derivation of the Hamiltonian function as $H = p_i q_i -L$, where $L$ is the Lagrangian function.
Dividing Eq.\ (\ref{Ec}) by $E$ and recalling that ${\bf \dot x}= {\bf p}/E$ and that $m/E = \sqrt{1- \left| {\bf \dot x} \right|^2}$, the relativistic invariance  of Eq.\ (\ref{H_Br}) is easily verified.
Eq.\ (\ref{H_Br}) has the same structure of the Dirac Hamiltonian. In particular the Dirac equation can be derived from it after the following additional replacements: ${\bf \dot x} \to {\bm \alpha}$, and $\sqrt{1- \left| {\bf \dot x} \right|^2} \to \beta$. The correspondence can be unequivocably proved calculating
the expectation values of the Dirac matrices  by using the solutions of the Dirac equation for a particle of definite momentum ${\bf p}$,
\begin{eqnarray}
	&&\langle {\bm \alpha} \rangle = {\bf p}/E = {\bf \dot x}\, ,\nonumber \\
	&&\langle {\beta} \rangle = m/E = \sqrt{1- \left| {\bf \dot x} \right|^2}\, .\nonumber
\end{eqnarray}
It may be surprising that the few-line derivation of the classical origin of Dirac equation  by Breit was scarcely exploited in the  literature on the interpretation of the Dirac equation.
Moreover, even more surprisingly, it is not described in textbooks on QFT at our knowledge (see e.g. \cite{weinberg,peskin,maggiore,mandl,hey}). Only ${\bm \alpha}$ is identified as the velocity operator from the Heisenberg equation 
$d{\bf \hat x}/dt = [{\bf \hat x}, \hat H]$ and/or from the current density operator \cite{Dirac,weinberg,peskin,hey}.
We believe that such a simple and profound result has been hidden by the relentless quest for explicit covariance.
Ordinary velocity is not considered in special relativity as a fundamental variable, it is not a 4-vector or part of it. Moreover it could be argued that Eq.\ {\ref{H_Br}} is not a proper classical Hamiltonian since it depends on velocity. 
Nevertheless it correctly  expresses the enegy of a  free relativistic particle,
$
E = {\bf \dot x} \cdot {\bf p}  + m\sqrt{1- \left| {\bf \dot x} \right|^2}
$. Moreover the Dirac Hamiltonian depends explicity on the velocity operator ${\bm \alpha}$, hence
one should not be surprised if its classical counterpart do depends on the velocity of the particle.

Dirac equation is generally presented in its explicit covariant form
\begin{equation}
	(i \gamma^\mu p_\mu - m) \psi =0\, ,
\end{equation}
with $\gamma^0= \beta$ and $\gamma^i=\gamma_0 \alpha_i$ ($i=1,3$).
In contrast to ${\bm \alpha}$ and $\beta$, the matrices $\gamma^\mu$
transform as a 4-vector under the application of the spinor representation of a Lorentz transformation: $\Lambda_{\frac{1}{2}}^{-1} \gamma^\mu \Lambda_{\frac{1}{2}} =\Lambda^{\mu}_{\nu}\gamma^{\nu}$,
where
\begin{equation}
	\Lambda_{\frac{1}{2}} = \exp({-\frac{i}{2} \omega_{\mu \nu} S_{1/2}^{\mu \nu}})\, 
\end{equation}
is the spinor representation of a Lorentz transformation, being $S_{1/2}^{\mu \nu} =(i/4)[\gamma_\mu, \gamma_\nu]$ the generators, and $\omega_{\mu \nu}$ an antisymmetrix matrix defining the specific transformation.
The  homogeneous Lorentz group denoted as $O(3,1)$ is the subgroup of Poincar$\acute{\text{e}}$ transformations describing rotations and boosts.
It is defined as the  group of  linear coordinate transformations.
$ x^{\mu} \to x^{'\mu} = \Lambda^{\mu}_{\nu}x^{\nu}$
which leave invariant the {\em proper time} interval $d \tau$.
Elements with det$\Lambda=1$  (called proper Lorentz transformations) form a subgroup denoted $SO(3,1)$.
In order to study the transformation properties of multiple-component quantum fields,
it is necessary to look for the finite-dimensional representations  of $SO(3,1)$ as  the spinor representation $\Lambda_{\frac{1}{2}}$. 
Although the $SO(3,1)$ algebra can be written as the algebra of $SU(2) \times SU(2)$, the group $SO(3,1)$ is non-compact.
Hence it has no faithful, finite-dimensional representations that are unitary despite the group $SU(2) \times SU(2)$ has.
The homogeneous Lorentz group, is thus the only group of relativistic QFT acting on multiple component quantum fields non-unitarily \cite{Fuchs}.
This rather surprising fact (at least apparently) conflicts with an important theorem proved by Wigner in 1931 (see Ref.\ \cite{weinberg} Chapter 2) which tells us that   any symmetry operation on quantum states must be induced by a unitary (or anti-unitary) transformation. The conflict is overcome, either by regarding the field not as a multicomponent quantum wavefunction but as a classical field \cite{peskin}, or by pointing out that the fundamental group is not the (homogeneous) Lorentz group but the
Poincar$\acute{\text{e}}$ group \cite{weinberg,maggiore}.
Independently on the point of view, one consequence is that the Hermitean conjugate $\psi^\dag$ of the (four-component) spinor field $\psi$ does not have the inverse transformation property of $\psi$ as requested by quantum mechanics. The solution  is to define $\bar \psi = \psi^\dag \gamma^0$ called the {\em Dirac conjugate} of $\psi$, being $\gamma^0$ the {\em time} Dirac matrix \cite{weinberg,peskin,maggiore,mandl,hey}.
The Lorentz-invariant Dirac Lagrangian can thus be written as
\begin{equation}
{\cal L}_{\text{Dirac}} =\bar \psi(i \gamma_\mu \partial_\mu - m) \psi\, .
\label{DiracL1}\end{equation}
It is quite curious that, by exploiting the relationship between the ${\bm \alpha}$ and ${\bm \gamma}$ matrices, it can be expressed without the need for the Dirac conjugate and in terms of the Dirac Hamiltonian and hence of the matrices ${\bf \alpha}$ and $\beta$:
\begin{equation}
{\cal L}_{\text{Dirac}} =\psi^\dag( i \partial_t - \hat H) \psi\, .
\label{DiracL2}\end{equation}

As discussed above, Breit found an expression for the classical Hamiltonian of a relativistic pointlike particle linear in the momentum and the mass parameter, which is the classical correspondent of Dirac's Hamiltonian \cite{Breit}. Eq.\ (\ref{H_Br}) also depends  on the ordinary velocity and on $1/(dt/d \tau) =\sqrt{1- \left| {\bf \dot x} \right|^2}$.
This correspondence suggests that these velocity variables and their simmetry properties play a fundamental role.
Let us consider an inertial reference frame $S$ including a clock placed at rest measuring the time $t$.
Let us consider a pointlike particle moving with a velocity ${\bf \dot x}$ relative to the inertial system $S$.
We will indicate by the vector ${\bf x}$ the position of the particle with respect to $S$.
Let us indicate by $\tau$ the time measured by a clock moving with the particle.
The proper time interval $d \tau$ is the time interval measured by a clock  fixed in the reference frame $S'$ which sees the particle at rest.

In this paper we investigate the very simple transformation properties of the ordinary velocity ${\bf \dot x}$ and of $1/(dt/d \tau)$.
In analogy to the  ordinary velocity components,  defined as $ \dot x_i \equiv d x_i/ dt$, we can regard
the fourth relevant variable as the proper time speed: $\dot \tau$.
Specifically, regarding the  time of the particle (like the position) as a function of time, we can invert obtaining $\dot \tau \equiv d \tau / dt = 1/(dt/d \tau)$. $\dot \tau$ describes the rate of ticking of particle's clock with respect that of the reference frame clock \cite{nota}.
Eq.\ (\ref{H_Br}) can thus be written  as
\begin{equation}
H = {\bf \dot x} \cdot {\bf p}  + \dot \tau\, m\, .
\label{E2}\end{equation}
We show that the  relativistic linear transformations, acting on the space of velocities rather than on coordinates,  display some relevant advanteges as compared to the homogenous Lorentz group: (i) they  are able to describe the spin additional degree of freedom of a pointlike particle yet at a classical  relativistic level; (ii)
they form a compact group hence with unitary finite-dimensional representations; (iii)
they describe antiparticles at a classical level as a direct consequence of rotation symmetry.
In addition we show that these linear transformations, acting on  rates of change with respect to the time coordinate $t$ of the reference frame,  attribute to  the latter a special role as required by quantum mechanics. Hence the proposed symmetry group  holds promise for a better reconciliation between relativity and quantum mechanics.

\section{S0(4) spacetime transformations}
The velocity of the particle  and  particle-time speed satisfy the following relationship,
\begin{equation}
 {\bf \dot x}^2 +  {\dot \tau}^2 = 1\, .
\label{epstein}\end{equation}
independently of the inertial reference frame.

Linear transformations which leave invariant the norm  in a four-dimensional (4D) Euclidean space (here defined by $(\dot x_1,\dot x_2,\dot x_3,\dot \tau)$\, ) constitute the  group SO(4). This group is not simple and has the same algebra of $SU(2) \times SU(2)$ as $SO(3,1)$ but (in contrast to $SO(3,1)$) is compact.
The kinematics of a relativisitc pointlike particle can be easily understood in terms of these variables. Eq.\ (\ref{epstein}) can be viewed as the norm of a unit 2D vector, which for later convenience we express as a 3D unit vector lying on  the ${\bf i}{\bf k}$ plane: ${\bf r} \equiv(\pm\left| {\bf \dot x} \right|,0,{\dot \tau})$.
Fig.\ 1 displays one such kinematic vector with positive components.
If the particle is at rest with respect to the reference frame, the vector lies on the ${\bf k}$ axis (${\dot \tau} = 1$) moreover  $\dot \tau$ decreases with particle-speed  increasing in the way predicted by special relativity (time dilation).
\begin{figure}[!ht]
\begin{center}
\resizebox{!}{4.1cm}{
\includegraphics{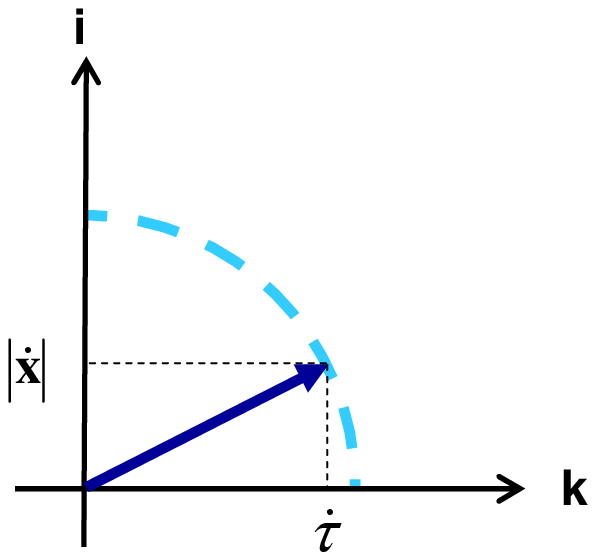}}\caption{Representation of the unit vector ${\bf r} \equiv(\left| {\bf \dot x} \right|,0,{\dot \tau})$}\label{fig1f}
\end{center}
\end{figure}
The unit vector ${\bf r}$, besides ${\dot \tau}$, is able only to describe the signed  modulus of the particle velocity $\pm\left| {\bf \dot x} \right|$. The particle velocity is actually a 3D vector and the direction of ${\bf \dot x}$ can be accounted for by one additional 3D unit vector ${\bf s}$ providing just the direction of ${\bf \dot x}$.
Hence the motion state of a particle can  be described by a specific couple of unit vectors ${\bf r}$ and ${\bf s}$:
\begin{eqnarray}
{\dot \tau} &=& {r}_3 \nonumber\\
{\bf \dot x}&=&  {r}_1\, {\bf s}\, .
\end{eqnarray}
Within this approach changes of the particle velocity
respect to an inertial frame can be accounted for by rotations of ${\bf r}$ in the kinematic plane  (changes of the  modulus) and rotations of ${\bf s}$ (changes of the  direction).
${\bf s}$ can be transformed according to arbitrary 3D rotations around an arbitrary 3D unit vector ${\bf n}$.
$s_i \to s_i' = [{R}_{\bf n}(\theta)]_{ij}\, s_j$ ($\theta$ labels the angle of rotation about ${\bf n}$) . Physical  kinematic states ${\bf r}$ admit only 2D rotations about the ${{\bf j}}$-axis: $r_i \to r_i' = [R'_{{\bf j}}(\phi)]_{ij}\, r_j$.
According to these rotation symmetries, states obtained rotating
${\bf r}$, and ${\bf s}$ should be considered as possible states. In particular
 ${\bf r}$ also describes
kinematic states in the second and third quadrant with ${\dot \tau} < 0$.
These states provide a classical description of antiparticles. This point will appear more evident after quantization.
From the point of view of classical (not quantum) special relativity, the description in terms of ${\bf s}$ and ${\bf r}$ appears to be redundant: a given velocity is described by two different states. For example a given velocity along direction ${\bf d}= {\bf \dot x}/ \left| {\bf \dot x} \right|$  can be described by the unit vectors
${\bf s}_\uparrow = {\bf d}$ and ${\bf r}_\uparrow = (\sin \theta ,0,\cos \theta)$ with $\theta = \arcsin \left| {\bf \dot x} \right|$, or equivalently by the unit vectors ${\bf s}_\downarrow =-{\bf d}$
and ${\bf r}_\downarrow = (\sin \theta' ,0,\cos \theta')$ with $\theta' = -\theta$.
Thus a given physical velocity and the corresponding proper-time speed are described by
two states. This twofold degeneracy recalls  quantum spin,  which Pauli  described in his paper on the exclusion principle \cite{Pauli}  as a {\em classical nondescribable two-valuedness}. This two-valuedness can be described in terms of the helicity variable $h = ({\bf s} \cdot {\bf p})/p$. 
In the following we will demonstrate that this classical twofold degeneracy is the classical correspondent of the helicty states determined by quantum spin. 
It is worth pointing out that, although the present approach describes a spin-like degree of freedom yet at a classical level, the interaction of a classical particle with the electromagnetic field does not appear to be affected by this additional degree of freedom, in contrast to what happens after quantization.
\begin{figure}[!ht]
\begin{center}
\resizebox{!}{4.5 cm}{
\includegraphics{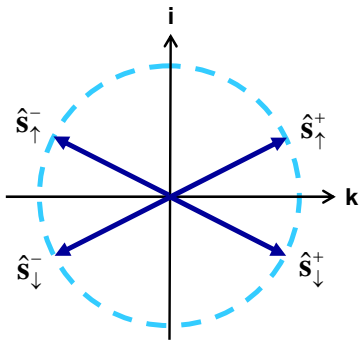}}
\caption{(Color online). Unit vectors ${\bf s}$ in the first and fourth quadrants describe spin up and spin down particles respectively. Unit vectors in the second and third quadrant describe spin up and spin down antiparticles.} \label{tintegrated ky=0}
\end{center}
\end{figure}
Figure\ 2 provides a clear geometric interpretation of the different kind of kinematic states: the first quadrant contains spin up particles, the second one spin up antiparticles, the third spin down antiparticles and the fourth quadrant spin down particles.
If we want to realize  these transformation properties on  quantum states, we have to find the representations of
the v-group acting on complex Hilbert spaces. The algebra of the v-group (as that of $SO(3,1)$) can be written as the algebra of $SU(2) \times SU(2)$.
As it is well known from non-relativistic quantum mechanics, the rotation group has finite representations
that are unitary with its generators represented by
angular momentum matrices. There is one irreducible representation $SU(N)$ for each finite dimension $N =2j +1$  with $j$ integer or semi-integer.
Being the v-group the tensor product of two independent rotations, the corresponding quantum generators will be a pair of independent angular momentum matrices.
Let us consider transformations on  elementry complex Hilbert spaces namely two level systems. In this case the generators are represented by pairs of Pauli matrices ${\bm \sigma}$ and ${\bm \rho}$ acting on two different 2D Hilbert spaces. These matrices have the same transformation properties of their classical counterparts ${\bf r}$ and ${\bf s}$. E.g.: 
\begin{equation}
D^{r \dag}_{{\bf j}}(\phi)\, \rho_i\, D^r_{{\bf j}}(\phi) = [R^r_{{\bf j}}(\phi)]_{ij}\, \rho_j\, ,
\end{equation}
being $D^r_{{\bf j}}(\phi) = \exp{\left( -i \rho_2\,  \phi/{2}\right)}$ .
As a consequence their expectation values transform as classical vectors. They are the quantum analog of classical unit vectors.
Thus, in order to combine rotation symmetry with quantum mechanics we can associate Pauli matrices to classical unit vectors. Applying this concept to the v-group above described we obtain:
\begin{eqnarray}
{\dot \tau} &=& {r}_3  \to \rho_3 \otimes I = \beta \nonumber\\
{\bf \dot x}&=&  {r}_1\, \hat{\bf s} \to \rho_1 \otimes {\bm \sigma} = {\bm \alpha}\, ,
\label{corr1}\end{eqnarray} being $I$ the identity operator in the $s$ space describing the direction of velocity.
It turns out that ${\bm \alpha}$ and $\beta$ are the well-known Dirac matrices in the standard representation.
${\bm \alpha}$ is the velocity vector operator, and Eq.\ (\ref{corr1}) allows us to identify $\beta$ as the proper-time speed operator.
Also higher order angular-momentum operators transform as classical vectors.
However it is worth noticing that expectation values of Pauli matrices have an additional unique property:
they obey the following relationship: $\sum_i \langle \sigma_i \rangle^2 =1$, implying
$\langle {\bm \alpha} \rangle^2 + \langle \beta \rangle^2=1$ (compare with Eq.\ (\ref{epstein})). This property ensures a closer adherence of quantum expectation values
to classical values. For example, working with a 3D complex Hilbert space ($j=1$), there would be quantum states displaying both space and proper time speeds equal to zero: $\langle J_3 \rangle = \langle J_1 \rangle = 0$ in clear contrast with the behaviour of classical relativistic particles. This observation attributes a unique  role to spin $1/2$ particles as the quantum correspondents of classical pointlike particles. The study of higher order ($N >2$) representations is left for future work, we expect that $j=1$ representations describe vector fields.

\section{Geometric description of Dirac solutions}
In order to better point out our correspondence principle for the Dirac equation, we start from Eq.\ (\ref{E2}) describing the energy of a classical free pointlike particle. 
It is formally a linear function of  momentum and depends on the  components of the two unit vectors ${\bm  r}$ and ${\bm  s}$: $E = {\bf p} \cdot {{\bf s}}\; r_1 + m\, r_3$.
If the symmetry properties of these unit vectors have any physical relevance, we should be able to quantize Eq.\ (\ref{E2}) by replacing the spatial and proper time velocities  with the corresponding quantum operators according to (\ref{corr1}). Performing also the usual operator replacements
$E \to i {\partial}_t$ and ${\bf p} \to -i {\partial}_{\bf x}$,
the Dirac equation is indeed obtained:
\begin{equation}
i {\partial}_t \psi(t,{\bf x}) = \left(-i {\bm \alpha} \cdot {\partial}_{\bf x} +  \beta m \right) \psi(t,{\bf x}) \equiv \hat H \psi(t,{\bf x})\, ,
\label{Dirac}\end{equation}
where $\psi$ is a four component wave function. 
This alternative derivation of the Dirac equation  based on the symmetry properties of kinematic rates
demonstrates that the symmetry properties of the spatial and proper time velocities play a deep role. 
\begin{figure}[!ht]
\begin{center}
\resizebox{!}{5 cm}{
\includegraphics{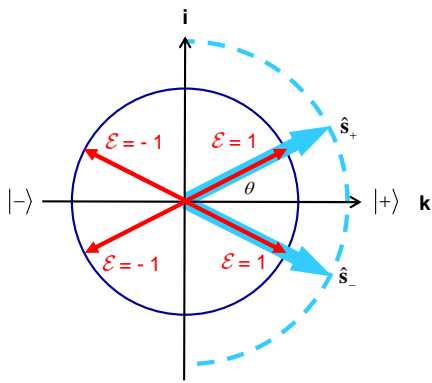}}
\caption{(Color online). Geometric representation of Dirac spinors on the ${\bf i}{\bf k}$ plane of the Bloch sphere. The larger arrows indicate the two possible kinematic unit vectors ${\bf  s}_{\pm}$ for a given momentum ${\bf p}$. The four thin arrows map  the corresponding four spinor eigenstates  lying on the ${\bf i}{\bf k}$ plane of the Bloch sphere. E.g. the thin arrow on the first quadrant map the $s$ state $\frac{1}{\sqrt{2}}(\cos \frac{\theta}{2} \left| + \right> +
\sin \frac{\theta}{2} \left| - \right>)$. The corresponding total spinor state is $\frac{1}{\sqrt{2}}\left| {\bf \tilde p},+ \right>_s(\cos \frac{\theta}{2} \left| + \right>_r +
\sin \frac{\theta}{2} \left| - \right>_r)$. }
\end{center}
\end{figure}
Let us look at the solutions of Dirac equation of the usual form
$
\psi = e^{-i E t}\, e^{i {\bf p} \cdot {\bf x}}  \left| \eta, \zeta \right>
$,
where $\eta$ and $\zeta$ are states belonging to the two 2D  Hilbert spaces (s and r) where $\sigma_i$ and $\rho_j$ act respectively. We assume that only ${\bf p}$ is determined and seek for $\eta$, $\zeta$, and $E$ solving the  eigenvalue problem obtained after inserting the above solution into Eq.\ (\ref{Dirac}):
\begin{equation}
{\cal H}  \left| \eta \right>_{s} \left| \zeta \right>_{r} = E \left| \eta \right>_{s} \left| \zeta \right>_{r}
\label{eigen}\end{equation}
with ${\cal H} = \rho_1\;  {\bm \sigma} \cdot {\bf p} + m\,  \rho_3$.
The  states in $s$ eigenstates of  ${\bm \sigma} \cdot {\bf \tilde p}$, where ${\bf \tilde p} = {\bf p}/p$:
\begin{equation}
{\bm \sigma} \cdot {\bf \tilde p} \left|{\bf \tilde p}, \pm \right>_{s} =
\pm \left|{\bf \tilde p}, \pm \right>_{s}\, ,
\end{equation}
 are eigenstates of ${\cal H}$. Inserting these eigenkets into Eq.\ (\ref{eigen}) and multiplying from the left for the same eigenbras, we are left with a pair of equations involving only states in $r$:
\begin{equation}
(\pm p \rho_1 + m\,  \rho_3) \left| \zeta \right>_{r} = E  \left| \zeta \right>_{r}\, ,
\end{equation}
which more compactly can be writen as

\begin{equation}
{\bf  s}_{\pm} \cdot {\bm \rho} \left| \zeta \right>_{r} = {\cal E} \left| \zeta \right>_{r}\, ,
\end{equation}
with ${\cal E} = E/\sqrt{p^2 + m^2}$ and
\begin{equation}
{\bf  s}_{\pm}= \frac{1}{ \sqrt{p^2 + m^2}} (\pm p,0,m) = (r_1,0,r_3) \, .
\end{equation}
The four Dirac solutions for a given ${\bf p}$ can thus be written as:
\[ \psi = e^{-i {\cal E} \sqrt{p^2+m^2}t} e^{i {\bf p} \cdot {\bf x}} \left|{\bf \hat p}, j \right>_{s} \left|{\bf  s}_{u}, {\cal E} \right>_{r}
\]
with ${\cal E},k = \pm 1$.
They have  a precise and simple geometric meaning that can be easily visualized on the Bloch sphere (Fig.\ 2). Positive energy solutions $\left|{\bf  s}_{\pm}, + \right>_{r}$ are given by  eigenstates of ${\bf  s}_{\pm} \cdot {\bm \rho}$ with ${\cal E} =1$ describing the system aligned along the unit vectors  ${\bf  s}_{\pm}$
in the first and fourth quadrant. Negative energy solutions $\left|{\bf  s}_{\pm}, + \right>_{r}$, corresponding to antiparticles, are given by  eigenstates of ${\bf  s}_{\pm} \cdot {\bm \rho}$ with ${\cal E} =-1$ and describe the system aligned along  directions in the second and third quadrant which are opposite to  ${\bf  s}_{\pm}$. The results displayed in Fig.\ 2 unequivocably show the correspondence between the classical and quantum descriptions of spin and antiparticles here proposed.
In particular solutions of the Dirac equation with negative energy (corresponding to antiparticles) display $\langle \beta \rangle < 0$, so the claim that classical states with
$\dot \tau =r_3 < 0$ are the classical correspondent of antiparticles is fully justified.
We finally observe that the present approach describes the transformation from an inertial reference frame to another by rotations, which after quantization imply the rotation on the Bloch sphere of the states $\left| \eta \right>_{s} \left| \zeta \right>_{r} \to \left| \eta' \right>_{s} \left| \zeta' \right>_{r}$. The correct coordinate dependence $\phi' = \phi (t',{\bf x}')$ of  the wave function corresponding to the rotated spinors $\phi (t',{\bf x}') \left| \eta' \right>_{s} \left| \zeta' \right>_{r}$ can be uniquely determined by solving the Dirac eigenvalue equation with $\phi (t',{\bf x}')$ as the unkown function. So doing (Lorentz) coordinate transformations can be recovered a posteriori in the spirit of a background-free theory.

\section{Discussion and Outlook}
We have presented new relativistic linear
transformations involving the ordinary velocity and  the proper-time rate of change with respect to the  time-coordinate of the frame of reference. According to them changes of the  velocity modulus
and  direction  can simply be  accounted for by rotations of two
independent unit vectors.  Dirac spinors just provide the quantum
description of these rotations.

Within this approach antiparticles
results from rotation symmetry. Moreover these transformations are
able to describe the spin additional degree of freedom of a
pointlike particle yet at a classical relativistic level. In
contrast to the homogeneous Lorentz group, they form a compact group
hence with unitary finite-dimensional representations as all other
symmetry groups in QFT. Hence the present approach is promising
towards a better reconciliation of relativity and quantum mechanics.
The approach here described provides a direct geometric visualization of Dirac spinors.  It sacrifies explicit covariance by making explicit  rotation symmetry which nevertheless is the fundamental symmetry on which the algebra of the Lorentz group is based.

A surprising feature of these results, requiring further
investigations, is that they have been obtained by regarding the
position-coordinate and proper-time  as functions of the
time-coordinate of the reference frame, thus attributing to  the
latter a special role as required by quantum mechanics. In so doing
these transformations put on the same footing  $d{\bf x}$ and $d
\tau$,  suggesting for the mass parameter $m$ the role of momentum
operator conjugate to $\tau$ as $-i {\bm \nabla}$ to ${\bf x}$. The
symmetric structure of Eq.\ (\ref{E2}) enforces this suggestion.
This would imply an internal time- energy uncertainty principle $\Delta \tau\, \Delta m$,
in agreement with the evidence of a gedanken experiment proposed by Aharonov and Rezni \cite{Ahranov}.

Further investigations are also required to understand the physical
meaning of the ${\bf j}$ axis in the $r$ space and hence of the
operator $\rho_2$ which is the generator of speed changes. It turns
out that all the kinematic states in special relativity are
described by unit vectors on the ${\bf i}{\bf k}$ plane and
consequently Dirac eigenstates are made of $r$ states on the
${\bf i}{\bf k}$ plane of the Bloch sphere. We envisage that
states outside the ${\bf i}{\bf k}$ plane play a role when
taking into account gravity.
\acknowledgments
We would like to thank B.\  Lucini and O.\ M.\ Marag\`{o} for helpful discussions and suggestions.


\begin{thebibliography}{50}


\bibitem{Pauli} W.\ Pauli, Z.\ Phys.\ {\bf 31}, 765 (1925).



\bibitem{CS1} Z.\ Grossmann and A.\ Peres, Phys.\ Rev.\ {\bf 132}, 2345 (1963).

\bibitem{CS2} A.\ O.\ Barut and N.\ Zanghi, Phys.\ Rev.\ Lett.\ {\bf 52}, 2009 (1984).

\bibitem{Newman} E.\ T.\ Newman, Phys.\ Rev.\ D {\bf 65}, 104005 (2002).


\bibitem{weinberg} S.\ Weinberg, {\em The Quantum Theory of Fields vol. 1} (Cambridge University Press, Cambridge, 1995).


\bibitem{Dirac}  P.\ A.\ Dirac {\em The principles of quantum mechanics} (Oxford University Press, New York, 1958).

\bibitem{Breit} G.\ Breit, Proc.\ Nat.\  Acad.\ Sci.\ {\bf 14}, 555 (1928).

\bibitem{Fuchs} J.\ Fuchs and C.\ Schweigert, {\em Symmetries, Lie Algebras and Representations}
(Cambridge University Press, Cambridge, 1995).




\bibitem{peskin}  M.\ E.\ Peskin and D.\ V.\ Schroeder {\em An ntroduction to quantum field theory} (Perseus Books, Reading, MA, 2000).

\bibitem{maggiore}  M.\ Maggiore {\em A modern introduction to quantum field theory} (Oxford University Press, New York, 2005).

\bibitem{mandl} F.\ Mandl and G.\ Shaw, {\em Quantum field theory} (Wiley, Chichester, 1984).

\bibitem{hey} I.\ Aitchison and A.\ J.\ Hey, {\em Gauge theories in particle physics, 2nd ed.} (Adam Hilger, Bristol, 1989)

\bibitem{nota} If someone dislikes regarding $\tau$ as a function of $t$, $\dot \tau$ can be defined as: $\dot \tau \equiv 1/(d t / d\tau)$.

\bibitem{Ahranov} Y.\  Aharonov, and B.\  Reznik, Phys.\ Rev.\ Lett.\ {\bf 84}, 1368 (2000).





















\end{thebibliography}
\end{document}